\documentstyle[12pt, aasms4]{article}

\begin{document}

\title{\hspace{0.5in} Discovery of $z=0.0912$ and $z=0.2212$
Damped Ly$\alpha$ Absorption 
Line Systems Toward the Quasar OI 363: Limits on the Nature of
Damped Ly$\alpha$ Galaxies\footnotemark}
\footnotetext{Based on  data obtained with the NASA/ESA
Hubble Space Telescope at the Space Telescope Science
Institute, which is operated by the Association of Universities
for Research in Astronomy, Inc., under NASA contract NAS5-26555, and
 on observations made with the KPNO WIYN telescope, National Optical
Astronomy Observatories, operated by AURA, Inc., under 
cooperative agreement with the National Science Foundation.}

\author{Sandhya M. Rao\altaffilmark{2}
 and David A. Turnshek\altaffilmark{3}
\affil{Department of Physics and Astronomy, University of Pittsburgh,
Pittsburgh, PA 15260}
\altaffiltext{2}{E-mail: rao@phyast.pitt.edu}
\altaffiltext{3}{E-mail: turnshek@vms.cis.pitt.edu}}

\begin{abstract}
The discovery of a $z_{abs}=0.0912$ damped Ly$\alpha$ absorption-line
system in the HST-FOS ultraviolet spectrum of the quasar OI$\;$363 (0738+313)
is reported.  This is the lowest redshift quasar damped Ly$\alpha$
system known. Its neutral hydrogen column density is $N$(\ion
H 1)$=1.5 \pm 0.2 \times10^{21}$ atoms cm$^{-2}$, which easily
exceeds the classical criterion for damped Ly$\alpha$ of $N$(\ion H
1)$\ge2\times10^{20}$ atoms cm$^{-2}$.  Remarkably, a $z_{abs}=0.2212$
damped system with $N$(\ion H 1)$=7.9 \pm 1.4 \times10^{20}$ atoms
cm$^{-2}$ has also been discovered in the same spectrum.

In the past, the standard paradigm for damped Ly$\alpha$ systems has
been that they arise in galactic or protogalactic \ion{H}{1} disks with low
impact parameters in luminous galaxies.  However, WIYN imaging
of the OI$\;$363 field shows that none of the galaxies visible in
the vicinity of the quasar is a luminous gas-rich spiral with low
impact parameter, either at $z= 0.0912$ or $z=0.2212$.
Thus, these damped systems are among the clearest examples
yet of cases that are inconsistent with the standard damped Ly$\alpha$ $-$ 
HI-disk paradigm.

\end{abstract}


\section{INTRODUCTION}

While it is widely recognized that QSO damped Ly$\alpha$
absorption-line systems are important probes of galaxy formation,
their utility is hampered by their rarity, especially in
the ultraviolet at low redshift. For example, only one damped
Ly$\alpha$ line was discovered in the {\it Hubble Space Telescope}
Quasar Absorption Line Key Project (Jannuzi et al. 1998) and, at the
present time, only two damped systems have been confirmed from a survey
of QSOs with the {\it International Ultraviolet Explorer} (Lanzetta, Wolfe
\& Turnshek 1995).  In the ultraviolet ($\lambda < 3220$ \AA),
Ly$\alpha$ absorption has redshifts $z_{abs} < 1.65$.
This corresponds to look-back times up to 77\% of the age of
the Universe for q$_0=0.5$.  Thus, it is particularly important that
low-redshift examples of damped Ly$\alpha$ systems be discovered and
studied. This will eventually allow the evolutionary links between 
\ion{H}{1} in
nearby galaxies (Rao \& Briggs 1993) and the high-redshift damped
systems found in optical surveys (Wolfe et al. 1986; Lanzetta et
al. 1991; Wolfe et al. 1995) to be explored.

We recently completed a highly successful HST Faint Object Spectrograph 
program to discover low-redshift ($z<1.65$) damped Ly$\alpha$ lines in QSO 
spectra (Turnshek 1997;
Rao \& Turnshek 1998).  Coupled with earlier HST archival
work (Rao, Turnshek \& Briggs 1995), we have studied the ultraviolet
Ly$\alpha$ absorption line in 88 \ion{Mg}{2} absorption-line systems.
This targeted search has greatly increased the number of known
low-redshift damped systems.  We currently have 12 damped systems
in our sample, including two that were discovered serendipitously 
in spectral regions without available \ion{Mg}{2} information.

In this $Letter$, we report our discovery of the lowest
redshift damped Ly$\alpha$ absorption-line system known ----- a
$z_{abs}=0.0912$ system with neutral hydrogen column density $N$(\ion
H 1)$\approx1.5\times 10^{21}$ atoms cm$^{-2}$ 
in the spectrum of the quasar OI$\;$363 (0738+313; $z_{em}=0.630$). 
This object has a known \ion{Mg}{2} absorption-line
system at $z_{abs}=0.2213$ (Boulade et al. 1987), but no evidence
for an absorption system at $z_{abs}=0.0912$ had ever been reported
prior to our survey.
We obtained a HST-FOS G160L/BL spectrum of this object to determine
the nature of the Ly$\alpha$ line associated with the 
$z_{abs}=0.2213$ system and coincidentally discovered the
damped system at $z_{abs}=0.0912$.  Remarkably, the slightly higher
redshift system is also damped with $N$(\ion H 1)$\approx7.9\times
10^{20}$ atoms cm$^{-2}$.

In \S2 we describe our observations and data analysis. The
observations include the HST-FOS spectrum of OI$\;$363 and an
R image of the field obtained in seeing of 0.55\arcsec\ with the WIYN 3.5-m
Telescope on Kitt Peak. We find that none of the galaxies visible in
the vicinity of the quasar are luminous gas-rich galaxies with low
impact parameters, either at $z\approx 0.0912$ or $z\approx 0.2212$.
In \S3 we discuss these results.

\section{OBSERVATIONS AND DATA ANALYSIS}

\subsection{HST-FOS Spectroscopy}

A one-orbit ultraviolet spectrum of OI$\;$363 was obtained on
1996 May 15 with the HST-FOS using the
G160L/BL grating-digicon combination and one of the 1.0-PAIR 
(0.86\arcsec\ square) apertures. The exposure time was 1520 seconds.
The pipeline-processed data were re-sampled to the original
dispersion of 1.72 \AA\ per pixel (quarter-diode); the re-sampled
spectrum is shown in Fig. 1 along with the $1\sigma$ error array.
A signal-to-noise ratio of $\approx 10$ per resolution element
(diode) was measured near 1400 \AA\ in the continuum adjacent to
the positions of the wide absorption lines seen in the spectrum.

The most recent calibration and data reduction procedures have been
applied to the data as recommended by the HST-FOS STScI analysis team.
Pixel numbers 900 through 1199 of the G160L/BL mode have zero
sensitivity to dispersed light. These were used to determine a
{\it wavelength-independent} correction for scattered light and
background (Kinney \& Bohlin 1993 ISR CAL/FOS-103). The average
flux value in the zero-sensitivity pixels was then subtracted from
the data during  pipeline processing.  In addition, the new and
improved Average Inverse Sensitivity method of flux calibration was
used in the  pipeline. The flux in the core of the wide absorption
line at $\approx$ 1330 \AA\ is approximately equal to the 1$\sigma$
uncertainty in the background. Thus, the marginal non-zero intensity of 
this wide absorption line may be attributed to the uncertainty in the
background measurement. However, the core of the wide absorption line
at $\approx$ 1486 \AA\ is  a  $\gtrsim 5\sigma$ deviation from zero
flux, and this is not expected to be due to uncorrected scattered
light, dark counts or red-leak (Ed Smith, FOS Instrument Scientist,
private communication).

Because neither of the wide absorption lines has zero flux at
its line center, we suspect that there is a background subtraction
problem in the FOS G160/BL spectrum (see also Boiss\'e et al. 1998).  
First, both absorption systems at redshifts 0.0912 and 0.2212
have now been found to be 21 cm absorbers (Lane et al. 1998a; b). 
For 21 cm absorption to appear the \ion{H}{1} column
density must always be significantly greater than $N$(\ion H 1)$=2 \times 10^{20}$ 
atoms cm$^{-2}$ (Briggs 1988), which is the classical limiting criterion defined
in surveys for damped Ly$\alpha$ lines (Wolfe et al. 1986).
Simulations at the FOS G160L/BL resolution
show that when 21 cm absorption
is present, the resulting damped Ly$\alpha$ line should be black, or
nearly black, at the line center.
Secondly, as discussed below, Voigt damping
profiles corresponding to column densities $N$(\ion H 1)$>2\times 10^{20}$ 
 cm$^{-2}$ are found
to be excellent fits to both wide absorption lines (see Figs. 2
and 3).  Thus, the different non-zero flux values at the cores of
the two lines suggest that a wavelength-dependent correction for the
background is needed.  Another possibility  is
that there is some other UV-bright object in the 0.86\arcsec\ FOS aperture
that is either unabsorbed or partially absorbed, but our WIYN image
of the field clearly indicates that if this is the case, then the object
is not visible in R.  To confirm the existence of very closely
spaced gravitationally lensed components would require better imaging, but
 we have dismissed this possibility for now.

Given that both absorption systems have also been detected in 
21 cm observations, the wide absorption lines are certainly due to 
Ly$\alpha$. Their equivalent widths  indicate that they lie
on the damping part of the curve-of-growth. Their 
\ion H 1 column densities were determined as follows:
a continuum level (CL1) 
was selected by eye using data points on either side of the line
near 1330\AA. Seventeen percent of CL1 was subtracted so that the
line center had zero flux, resulting in a new continuum (CL2).
The spectrum was then normalized by CL2. Fig. 2 shows a Voigt
profile fit to the Ly$\alpha$ line for $z_{abs}=0.0937$ and $N$(\ion
H 1)$=1.5\times 10^{21}$  cm$^{-2}$. Similarly, for the line
near 1486\AA, 32\% of the original continuum was subtracted so
that the line center would have zero flux and the spectrum was 
re-normalized. Fig. 3 shows a
Voigt profile fit to this Ly$\alpha$ line for $z_{abs}=0.2224$ and
$N$(\ion H 1)$=7.9\times 10^{20}$ cm$^{-2}$.  The parameters that
best fit the data were determined by minimizing the least squares
difference between the data and the fit within 12\AA\ on either
side of the line near  1330\AA\ and within 17\AA\ on either side
of the line near 1486\AA.  The largest uncertainty in the
measured column densities comes from continuum placement.
An estimate of the uncertainty was determined by placing new
continua at levels corresponding to $CL2 \pm1\sigma$ 
with the $\sigma$ offset given by the error array. The
spectrum was then re-normalized using the new continua, and the
best fit damping profiles were re-determined.  The resolution (FWHM)
of the observations is approximately one diode, which is 6.6\AA.
The data were obtained with quarter-substepping so that each pixel
is $\approx$ 1.7\AA.  This suggests that the redshifts of the
damped Ly$\alpha$ lines in our G160L/BL spectrum can be determined to an
accuracy of $z\approx 0.001$. The associated 21 cm absorption lines
have redshifts 0.0912 and 0.2212, which differ by 2.5$\sigma$ 
and 1.2$\sigma$ from the damped line redshifts, respectively.   Since the 
narrow 21 cm line gives a highly accurate measurement of the 
redshift, the 21 cm measurements are reported as the redshifts of the 
two systems without error.   In summary, the
lower redshift damped Ly$\alpha$ line has $z_{abs}=0.0912$
and $N$(\ion H 1)$=(1.5\pm0.2)\times 10^{21}$ cm$^{-2}$, while the
 higher redshift line has $z_{abs}=0.2212$ and $N$(\ion H
1)$=(7.9\pm1.4)\times 10^{20}$ cm$^{-2}$.

\subsection{Imaging}

Images of the OI$\;$363 field were obtained for us by
the KPNO-WIYN Queue team on 1 October 1997 in the Harris R
filter using a Tektronics $2048\times2048$ CCD.    
The total exposure time was 2700 seconds.
The image in Fig. 4 was obtained by bias-subtracting, flat-fielding,
 registering and combining the original frames 
in the standard way.  The image scale
is 0.195\arcsec\ per pixel and the seeing was measured to be 
0.55\arcsec.  The $1\sigma$ surface brightness limit  is 
$\mu(R) = 25.4$ magnitudes arcsec$^{-2}$, while the point source 
limiting magnitude is $m_R = 25.8$. 

In Table 1, we summarize the positions and R-band magnitudes of 
resolved objects (presumably all galaxies) detected within
40\arcsec\ of the quasar and brighter than $R=24.0$. 
Objects whose radial light distributions had
FWHMs greater than 3.1 pixels, or 0.60\arcsec,  were taken to be  
resolved objects. The positions and R-band magnitudes of
all unresolved
point sources (presumably stars) that were detected within 40\arcsec\ of 
the quasar are listed in a footnote to Table 1. The cores of the 
quasar and the objects labeled S5, S10, and S11 are saturated.
Since the WIYN frame was 
uncalibrated, we assumed the magnitude of the star labeled S7 
to be $m_R=18.3$, as measured by Drinkwater et al. (1993) under
photometric conditions (see below). We then determined
the magnitudes of all the other objects by measuring
their intensities relative to S7.    
Only one object has a measured redshift. The spectrum of the galaxy G11 was 
obtained for us by Kathy Romer; its redshift is 0.06.
It is of  interest to consider the possible absolute magnitudes
and impact parameters of the other galaxies along the line of sight to
OI$\;$363 by assuming that they are at one of the two relevant
absorption redshifts.  This is done in Table 1. 
We also give the most likely morphological
type for each galaxy as determined either directly from the WIYN
image or from its absolute magnitude, as in the case of dwarfs.
The implications of these results are discussed in \S3.

In addition to our WIYN image, other imaging observations of the
OI$\;$363 field have been discussed in the literature.  
Le Brun et al. (1993) detected three galaxies  with apparent magnitudes
$r>24.5$ within 5.5\arcsec\ of the quasar. We detect a marginally significant 
extended feature to the
south-west of the quasar that coincides with one of these detections.
 We measured 20 counts per pixel above the local background,
i.e., a surface brightness of
25.3 magnitudes per square arcsec ($\approx$ the $1\sigma$ limit of our image),
 at the position marked with a cross in Fig. 4.  Le Brun et al. 
suggest that the three galaxies are associated with the quasar
and that the galaxy labeled G1 in Table 1 is the one most likely 
to give rise to the $z_{abs}=0.2213$ system. However,
this conclusion was reached before it was realized that there is
a system at $z_{abs}=0.0912$ and before both systems were recognized
to be damped Ly$\alpha$ absorbers.
Hutchings \& Neff (1990), Drinkwater et al. (1993), and Kirhakos et al. 
(1994) have also obtained images of OI$\;$363 and the surrounding field. 
Although
the data of Drinkwater et al. (1993) were obtained under photometric 
conditions, their seeing was only 2.8\arcsec. As a result, they identified
galaxy G1 as a star. They also detected two galaxies with  reported positions
near G2 with magnitudes brighter than our detection limit. These are
not visible in our WIYN image. 

\section{DISCUSSION}

The two damped Ly$\alpha$ systems described in this $Letter$ provide
an excellent opportunity to determine the nature of damped Ly$\alpha$
galaxies. It is interesting to note that none of the objects listed
in Table 1 conform to the standard paradigm that damped Ly$\alpha$
galaxies are  $M_{R}^* \approx -21$ spiral disks (Wolfe et al. 1986;
Wolfe 1988) with small impact parameters ($b<20h^{-1}_{75}$ kpc,
Steidel 1995), where we take $B-R=1$.  The usual convention is that
this $M^*$  characterizes the  Schechter luminosity function of all
galaxy types combined (see Fig. 2 in Rao \& Briggs 1993).  However,
the luminosity function of spiral galaxies is best described by a
Gaussian function with a peak occurring at $M_R^*\approx -18$ (see
Fig. 1 in Rao \& Briggs 1993).  Interestingly, the fact that brighter
spirals have larger \ion{H}{1} disks pushes the peak in the
interception-probability distribution, which is the product of number
density and \ion{H}{1} cross-section, to higher luminosities, i.e., to
$M_R \approx -21$ (see Fig. 5).

If any one of the faint galaxies reported by Le Brun et al. within
$\approx$ 5\arcsec\ of the quasar is the absorber, then it would be an
extremely faint dwarf galaxy at $z=0.0912$, with\\ $M_r$
$\gtrsim-13.5$. On the other hand, the  low-surface-brightness
feature, marked by a cross in Fig. 4, might be part of a galaxy
that lies directly in front of the quasar (in this case for $z=0.09$,
$M_R>-18$ is still likely).  Its existence needs to be confirmed.
The candidate G1 is a likely absorber at $z=0.0912$, with absolute
magnitude $M_R=-16.9$ ($M_B=-15.9$), $\approx 4$ magnitudes fainter
than the peak in the interception probability distribution.  It is a
better candidate at $z=0.2212$, where it would be brighter and still
at a reasonable impact parameter from the quasar. Galaxies G2
through G9 would be dwarfs at both redshifts and, given the high
column densities of the two damped Ly$\alpha$ lines,  make highly
unlikely absorbers at the listed impact parameters. The galaxy G10
appears to be an early type galaxy and, though reasonably bright, is
not expected to contain much gas, especially at large impact
parameters.  The galaxy G11 is the only large spiral in the field
(see the Fig. 4 caption); but its redshift is $z=0.06$ (\S2.2).
Thus, one could conclude that G1 is the only galaxy that is a
reasonable candidate at either absorption redshift.

The absence of luminous spiral galaxy candidates at both damped
absorption redshifts represents the most illustrative examples yet of
inconsistencies with the standard HI-disk paradigm for damped
Ly$\alpha$ systems. Others have discussed similar findings in
systems of higher redshift (Steidel et al. 1994; Steidel et al. 1997;
Le Brun et al. 1997), but our upper limits on the luminosity of
candidate damped Ly$\alpha$ galaxies are about an order of magnitude
lower than previous results.  Clearly, not all damped Ly$\alpha$
absorbers are luminous spirals.

It has been known for some time that $\Omega_{gas}$ at high-redshift
is comparable to the present-day luminous (stellar) cosmological mass
density (Rao \& Briggs 1993).  Thus, the high-redshift damped systems
are possibly the progenitors of all present-day galaxies, not just
luminous spiral HI-disks.  The recent work of Khersonsky \& Turnshek
(1996), Pettini et al. (1997), Rao \& Turnshek (1998), and references
therein also reveal problems with this scenario.  In fact, the
results of our recent survey for damped Ly$\alpha$ absorption at
$z_{abs}<1.65$ indicate that the cosmological mass density of neutral gas,
$\Omega_{gas} = 1.3\Omega_{HI}$, at $z \approx 1$ is also
comparable to the luminous mass density at the current epoch (Rao \&
Turnshek 1998).  One solution might be the existence of  many
unrecognized low-surface-brightness (LSB) galaxies or dwarf galaxies
at low redshift. However, Briggs (1997a; b) and Zwaan et al. (1997) have
concluded that the present-day population of LSB galaxies contributes
little to the \ion{H}{1} mass density at $z=0$.  They appear to be
drawn from the same population as the ``normal'' high-surface-brightness 
(HSB) galaxies that are found in optical surveys. Moreover, Bothun et al. 
(1997) have shown that the average \ion{H}{1} column density in LSB
galaxies is lower than that in HSB galaxies.  Rao \& Briggs (1993)
used {\it available data} to show that the local \ion{H}{1} mass
density and cross-section is dominated by the large spirals.  Thus,
based on the present imaging results, we conclude that the 
nature of the damped Ly$\alpha$ galaxies is not yet fully understood.

\noindent {\bf ACKNOWLEDGEMENTS.} We thank Ed Smith and Alex Storrs of
the HST-FOS team for help and advice with FOS data reduction. We also
thank Diane Harmer and Paul Smith for their assistance in obtaining
the WIYN data, and Kathy Romer for kindly obtaining the spectrum of
the $z=0.06$ galaxy (G11) in the field of the quasar OI$\;$363 with the
KPNO 4m.  This observation allowed us to exclude that galaxy as a
possible site for the absorption.  We especially thank Eric Monier
for his assistance with many aspects of this work.

\newpage

\figcaption[OI363.eps]{HST-FOS G160L/BL spectrum of the QSO OI363 with damped Ly$\alpha$
absorption lines at 1330\AA\ and 1486\AA. The $1\sigma$ error array is also
shown.}

\figcaption[z.0937fit.eps]{Theoretical Voigt profile with redshift $z=0.0937$ and 
neutral hydrogen column density $N(HI)=1.5\times 10^{21}$ atoms cm$^{-2}$
fit to the absorption line at 1330\AA. The normalized continuum and zero
levels are shown.}

\figcaption[z.2224fit.eps]{Theoretical Voigt profile with redshift $z=0.2224$ and 
neutral hydrogen column density $N(HI)=7.9\times 10^{20}$ atoms cm$^{-2}$
fit to the absorption line at 1486\AA. The normalized continuum and zero
levels are shown.}

\figcaption[q0738_Rlabels.ps]{A 1.5\arcmin\  $\times$ 1.5\arcmin\ field  of the 
WIYN image around the quasar OI 363. North is up and East is to the left.
The plate scale is 0.195\arcsec\ per pixel. The redshift of the galaxy marked
G11 is $z=0.06$; the expected appearance of a luminous gas-rich spiral at 
$z=0.09$ is qualitatively similar to G11.}

\figcaption[cs.ps]{The contribution of spiral and irregular galaxies to the 
\ion{H}{1} absorption cross-section for $N(HI) > 10^{19}$ cm$^{-2}$
at the current epoch assuming a Gaussian
luminosity function for spirals and a Schechter luminosity function
for irregulars (see Rao \& Briggs 1993). The peak in the spiral
distribution function occurs at $M_B\approx -20$, or $M_R\approx-21$
assuming $B-R=1$.  See Rao (1994) for details of the luminosity$-$diameter relations.}

\clearpage

\begin{table}
\begin{center}
\begin{tabular}{lrrrcccrcc} 
\multicolumn{10}{c}{Table 1}\\
\multicolumn{10}{c} {Extended Sources within 40\arcsec\ of the Quasar\tablenotemark{a}}\\
\tableline
Object & $\Delta\alpha(\arcsec)$&  $\Delta\delta(\arcsec)$ & 
$\Delta\theta(\arcsec)$ & m$_R$\tablenotemark{b} &   b\tablenotemark{c} (kpc)& 
M${_R}$\tablenotemark{c} &  b\tablenotemark{c} (kpc) &  
M${_R}$\tablenotemark{c} & Morphology\\
 &  &  &  &  & \multicolumn{2}{c}{(z=0.0912)} & \multicolumn{2}{c}{(z=0.2212)} & \\
%(1) & (2) & (3) & (4) & (5) & (6) & (7) & (8) & (9) \\
\tableline
G1 & 2.0 & $-$5.3 & 5.7 & 20.8 & 8.9 & $-$16.9 & 17.7 & $-$18.8 & ?\\
G2 & $-$10.9 & 3.1 & 11.3 & 23.4 & 17.7 & $-$14.3 & 35.0 & $-$16.0 &  dw\\
G3 & 7.2 & $-$9.0 & 11.5 & 23.7 & 18.0 & $-$14.0 & 35.7 & $-$15.9 &  dw\\
G4 & $-$10.3 & 8.8 & 13.5 & 22.5 & 21.1 & $-$15.2 & 42.2 & $-$17.1 &  dw\\
G5 & 8.4 & $-$11.3 & 14.1 & 24.0 & 22.1  & $-$13.7 & 43.7 &  $-$15.6 & dw\\
G6 & 9.8 & 12.1 & 15.5 & 22.0 & 24.3 & $-$15.7 & 48.1 & $-$17.6 & dw \\
G7 & $-$2.1 & 17.6 & 17.7 & 22.4 & 27.7 & $-$15.3 & 54.9 & $-$17.2 &  dw\\
G8 & $-$18.9 & 10.9 & 21.8 & 23.3 & 34.1 & $-$14.4 & 67.6 & $-$16.3 &  dw\\
G9 & 6.8 & 21.1 & 22.1 & 23.9 & 34.6 & $-$13.8 & 68.5 & $-$15.7 & dw\\
G10 & $-$26.5 & $-$9.2 & 28.1 & 19.8 & 44.0 & $-$17.9 & 87.1 & $-$19.8 & early type\\
G11\tablenotemark{d} & 31.4 & 1.2 & 31.4 & 17.1 &  &  &  &  & spiral\\
G12 & $-$25.4 & 18.9 & 31.6 & 22.2 & 49.5 & $-$15.5 & 98.0 & $-$17.4 &  dw\\
G13 & $-$30.4 & 11.1 & 32.4 & 22.2 & 50.7 & $-$15.5 & 100.4 & $-$17.4 & dw\\
G14 & $-$7.6 & 33.9 & 34.8 & 22.5 & 54.5 & $-$15.2 & 107.9 & $-$17.1 & dw\\
G15 & $-$31.8 & 17.2 & 36.2 & 22.0 & 56.7 & $-$15.7 & 112.2 & $-$17.6 & dw
\end{tabular}
\end{center}
\tablenotetext{a}{Point sources (Object, $\Delta\alpha(\arcsec)$, $\Delta\delta(\arcsec)$, m$_R$):  QSO, 0.0, 0.0, sat; S1, 2.0, 1.5, 20.8;
S2, $-$7.2, 5.3, 22.6;
S3, $-$0.9, 11.5, 20.5;
S4, $-$8.0, 9.8, 20.5;
S5, $-$12.8, 6.0, sat;
S6, $-$17.9, 3.3, 23.4;
S7, 3.3, $-$19.9, 18.3;
S8, $-$17.9, 14.2, 23.4;
S9, 11.1, 21.7, 21.3;
S10, 5.3, 24.2, sat;
S11, $-$34.7, $-$11.3, sat.}
\tablenotetext{b}{R-band magnitude of S7 from Drinkwater et al. (1993).}
\tablenotetext{c}{H$_0$ = 75 km s$^{-1}$ Mpc$^{-1}$, q$_0$=0.5.}
\tablenotetext{d}{The redshift of galaxy G11 is $z=0.06$.}
\end{table}

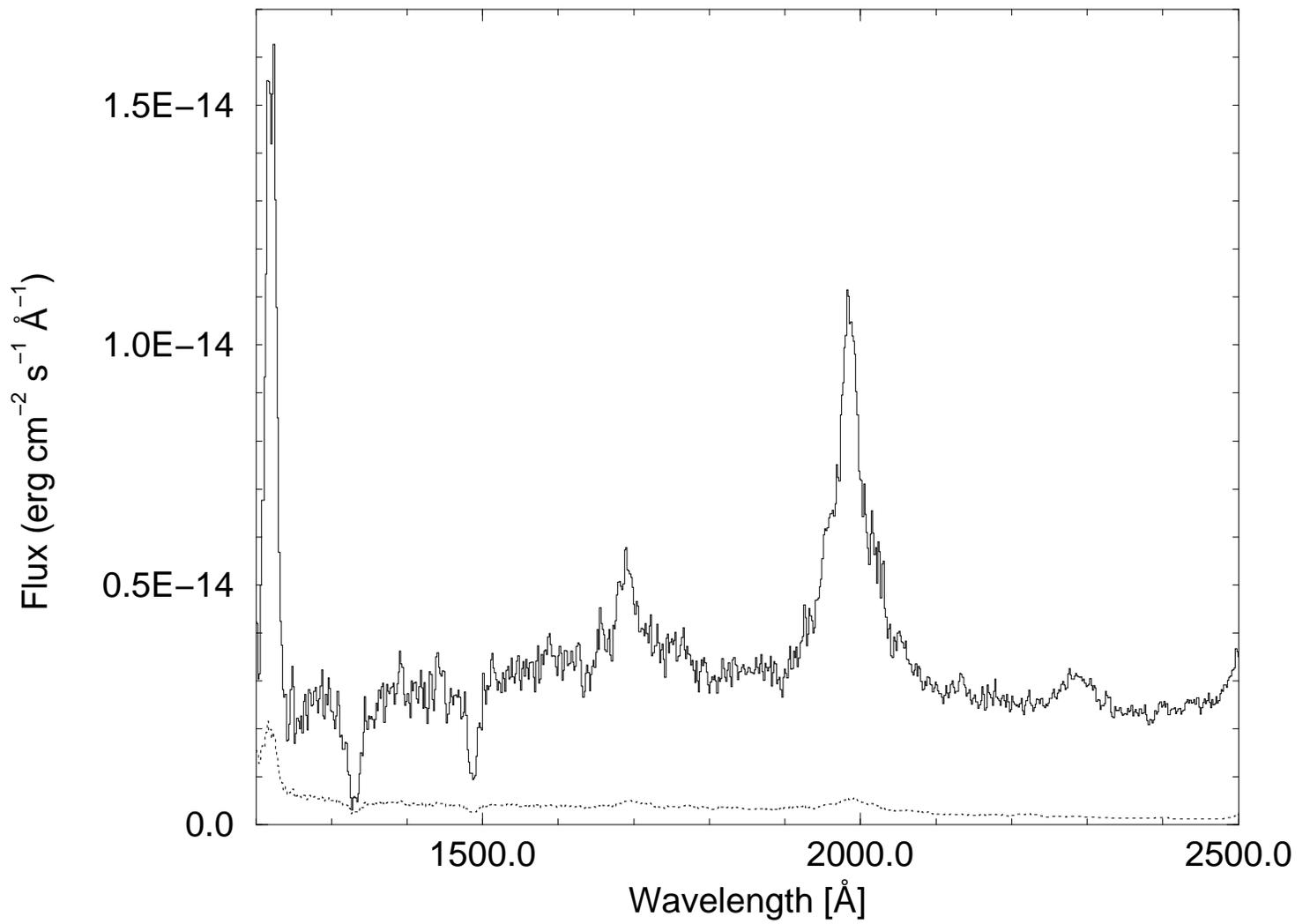
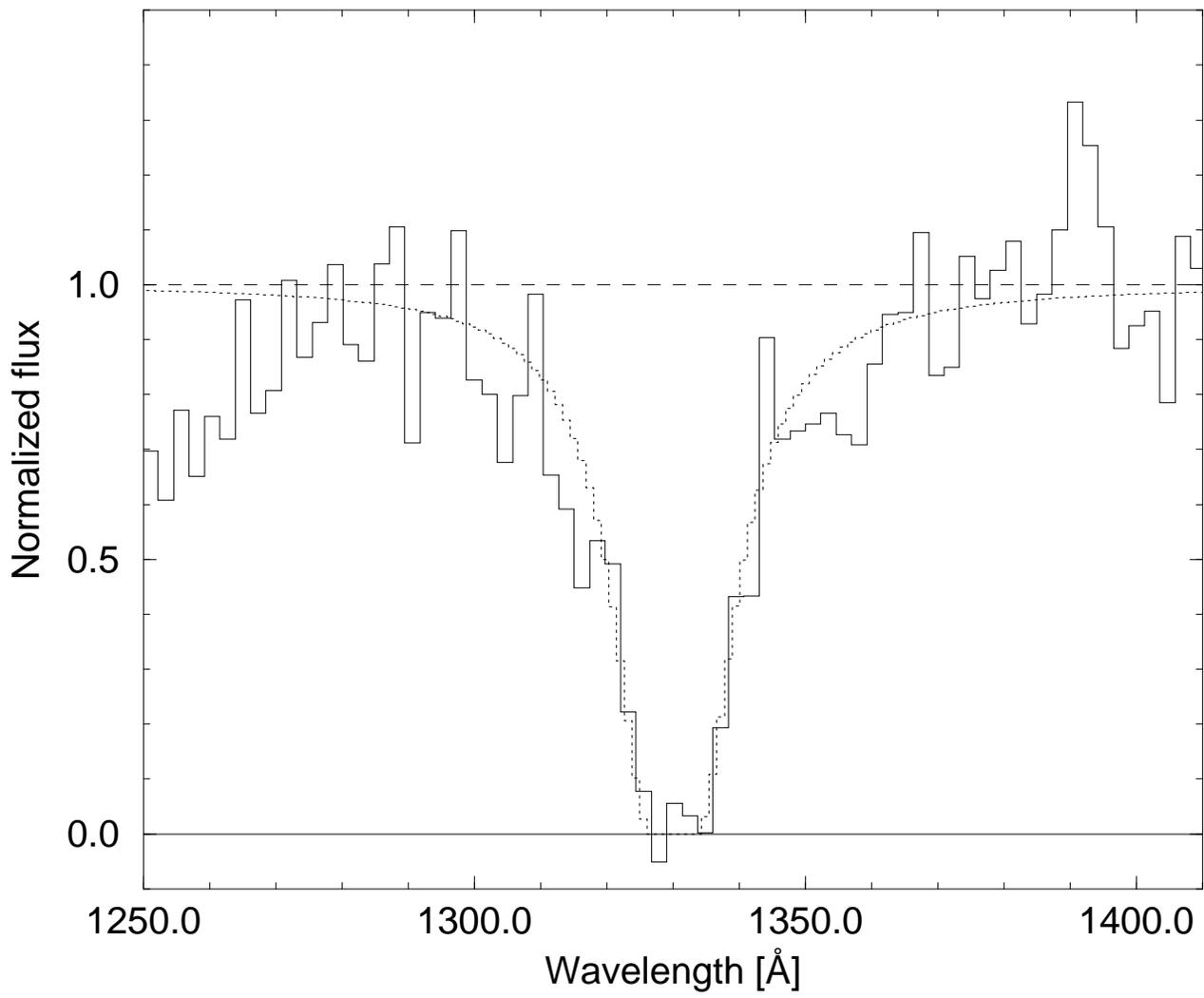
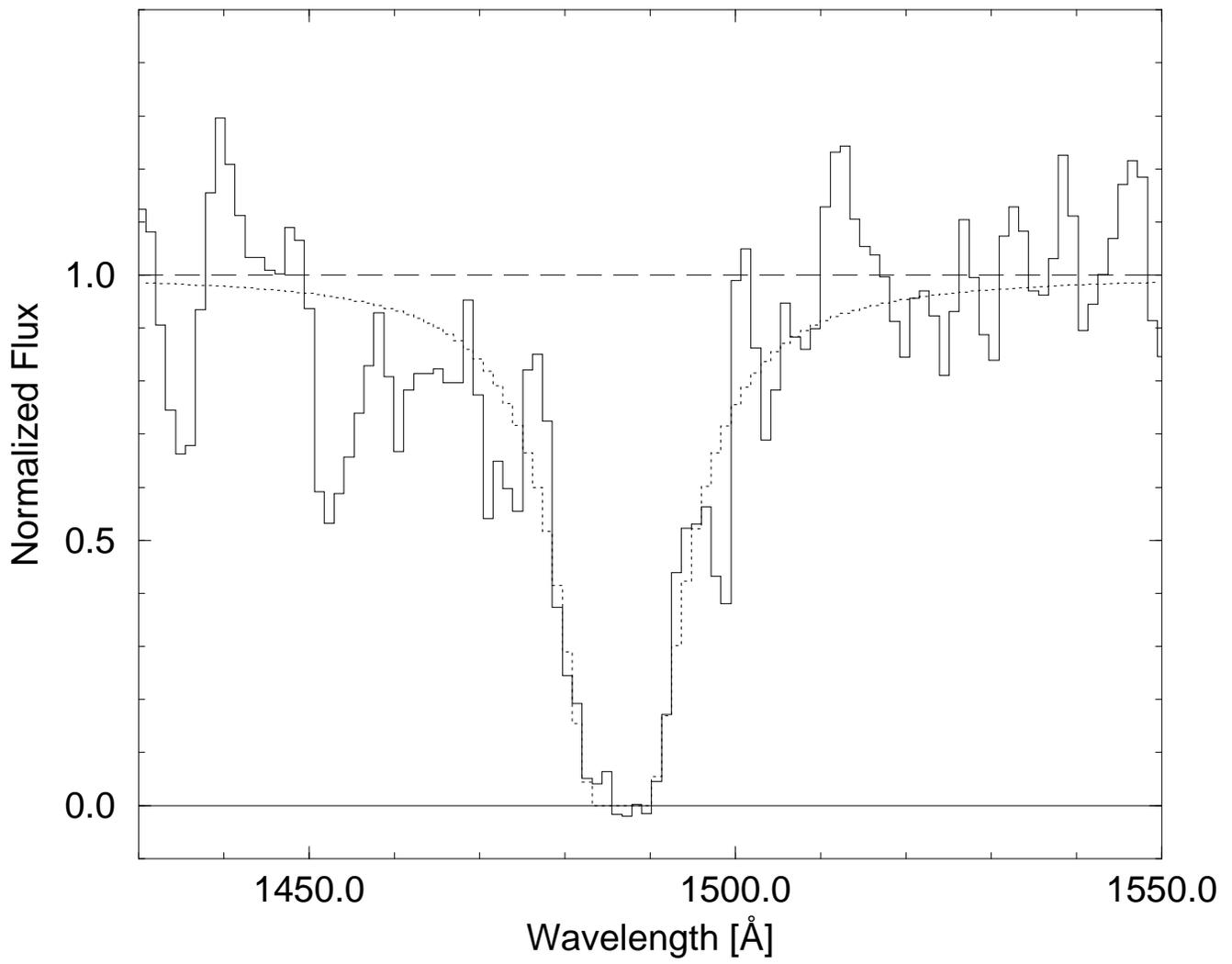
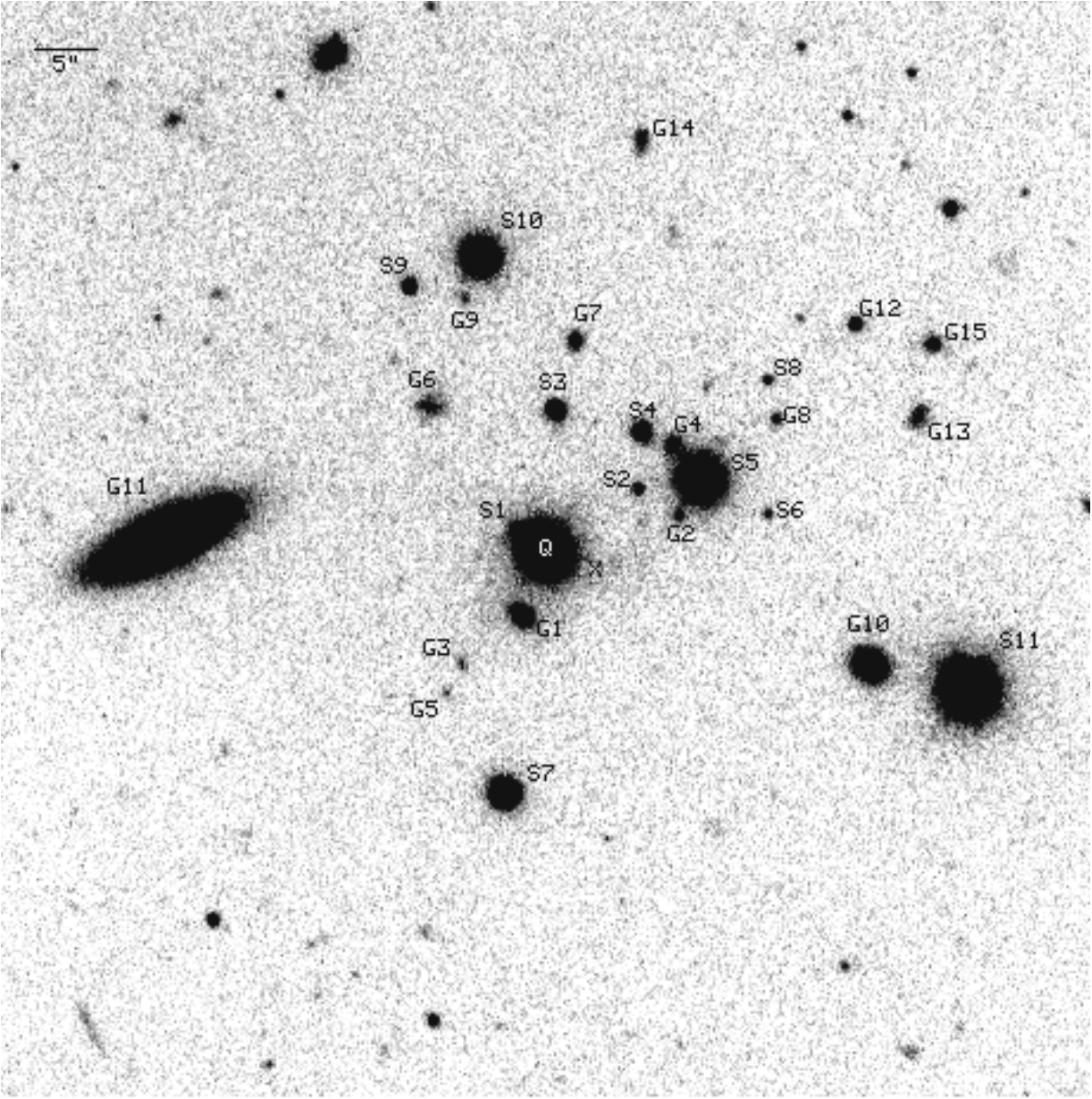
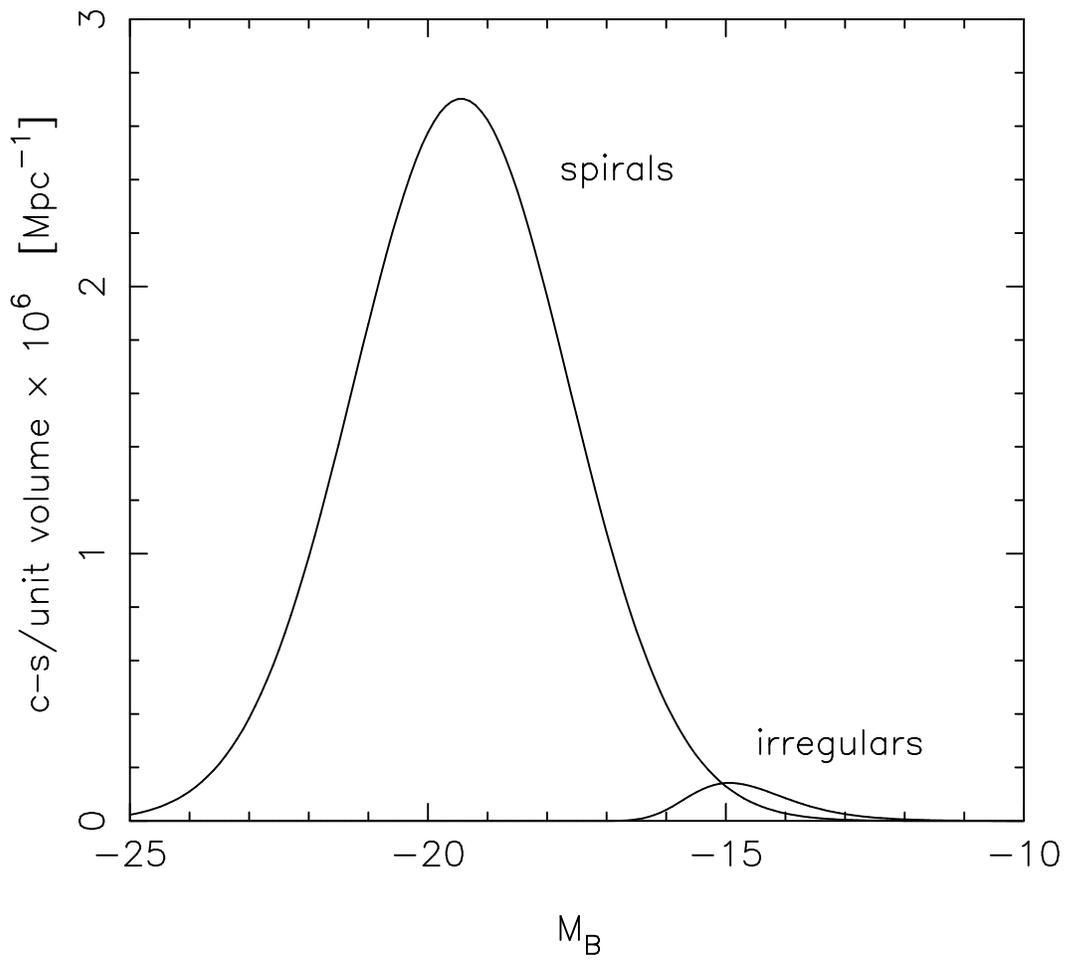
\end{document}